\documentclass[twoside,onecolumn,a4paper]{article}
\usepackage{amsmath, amsthm, amssymb}
\usepackage[super,sort&compress,comma]{natbib} 
\usepackage{times}

\usepackage{graphicx} %eps figures can be used instead
\usepackage{hyperref}

\begin{document}

\thispagestyle{plain}
\renewcommand{\thefootnote}{\fnsymbol{footnote}}
\renewcommand\footnoterule{\vspace*{1pt}% 
\hrule width 3.4in height 0.4pt \vspace*{5pt}} 
\setcounter{secnumdepth}{5}

\makeatletter 
\def\subsubsection{\@startsection{subsubsection}{3}{10pt}{-1.25ex plus -1ex minus -.1ex}{0ex plus 0ex}{\normalsize\bf}} 
\def\paragraph{\@startsection{paragraph}{4}{10pt}{-1.25ex plus -1ex minus -.1ex}{0ex plus 0ex}{\normalsize\textit}} 
\renewcommand\@biblabel[1]{#1}            
\renewcommand\@makefntext[1]% 
{\noindent\makebox[0pt][r]{\@thefnmark\,}#1}
\makeatother 
\renewcommand{\figurename}{\small{Fig.}~}

\setlength{\arrayrulewidth}{1pt}
\setlength{\columnsep}{6.5mm}
\setlength\bibsep{1pt}

\noindent\LARGE{\textbf{Water formation at low temperatures by surface O$_2$ hydrogenation I: characterization of ice penetration}}
\vspace{0.6cm}

\noindent\large{\textbf{S.~Ioppolo,$^{\ast}$\textit{$^{a}$}
H.~M.~Cuppen,\textit{$^{a,b}$} C.~Romanzin,\textit{$^{a\dag}$}
E.~F.~van Dishoeck\textit{$^{b,c}$} and
H.~Linnartz\textit{$^{a}$}}}\vspace{0.5cm}
%Please note that \ast indicates the corresponding author(s) but no footnote text is required. 

%\noindent\textit{\small{\textbf{Received Xth XXXXXXXXXX 20XX, Accepted Xth XXXXXXXXX 20XX\newline
%First published on the web Xth XXXXXXXXXX 200X}}}

\noindent \textbf{\small{DOI: 10.1039/C0CP00250J}}
\vspace{0.6cm}
%Please do not change this text.

\noindent \normalsize{Water is the main component of interstellar
ice mantles, is abundant in the solar system and is a crucial
ingredient for life. The formation of this molecule in the
interstellar medium cannot be explained by gas phase chemistry only
and its surface hydrogenation formation routes at low temperatures
(O, O$_2$, O$_3$ channels) are still unclear and most likely
incomplete. In a previous paper we discussed an unexpected
zeroth-order H$_{2}$O production behavior in O$_2$ ice hydrogenation
experiments compared to the first-order H$_2$CO and CH$_3$OH
production behavior found in former studies on hydrogenation of CO
ice. In this paper we experimentally investigate in detail how the
structure of O$_{2}$ ice leads to this rare behavior in reaction
order and production yield. In our experiments H atoms are added to
a thick O$_{2}$ ice under fully controlled conditions, while the
changes are followed by means of Reflection Absorption InfraRed
Spectroscopy (RAIRS). The H-atom penetration mechanism is
systematically studied by varying the temperature, thickness and
structure of the O$_{2}$ ice. We conclude that the competition
between reaction and diffusion of the H atoms into the O$_{2}$ ice
explains the unexpected H$_{2}$O and H$_{2}$O$_{2}$ formation
behavior. In addition, we show that the proposed O$_2$ hydrogenation
scheme is incomplete, suggesting that additional surface reactions
should be considered. Indeed, the detection of newly formed O$_{3}$
in the ice upon H-atom exposure proves that the O$_{2}$ channel is
not an isolated route. Furthermore, the addition of H$_{2}$
molecules is found not to have a measurable effect on the O$_{2}$
reaction channel.} \vspace{0.5cm}

\includegraphics[width=.44\textwidth]{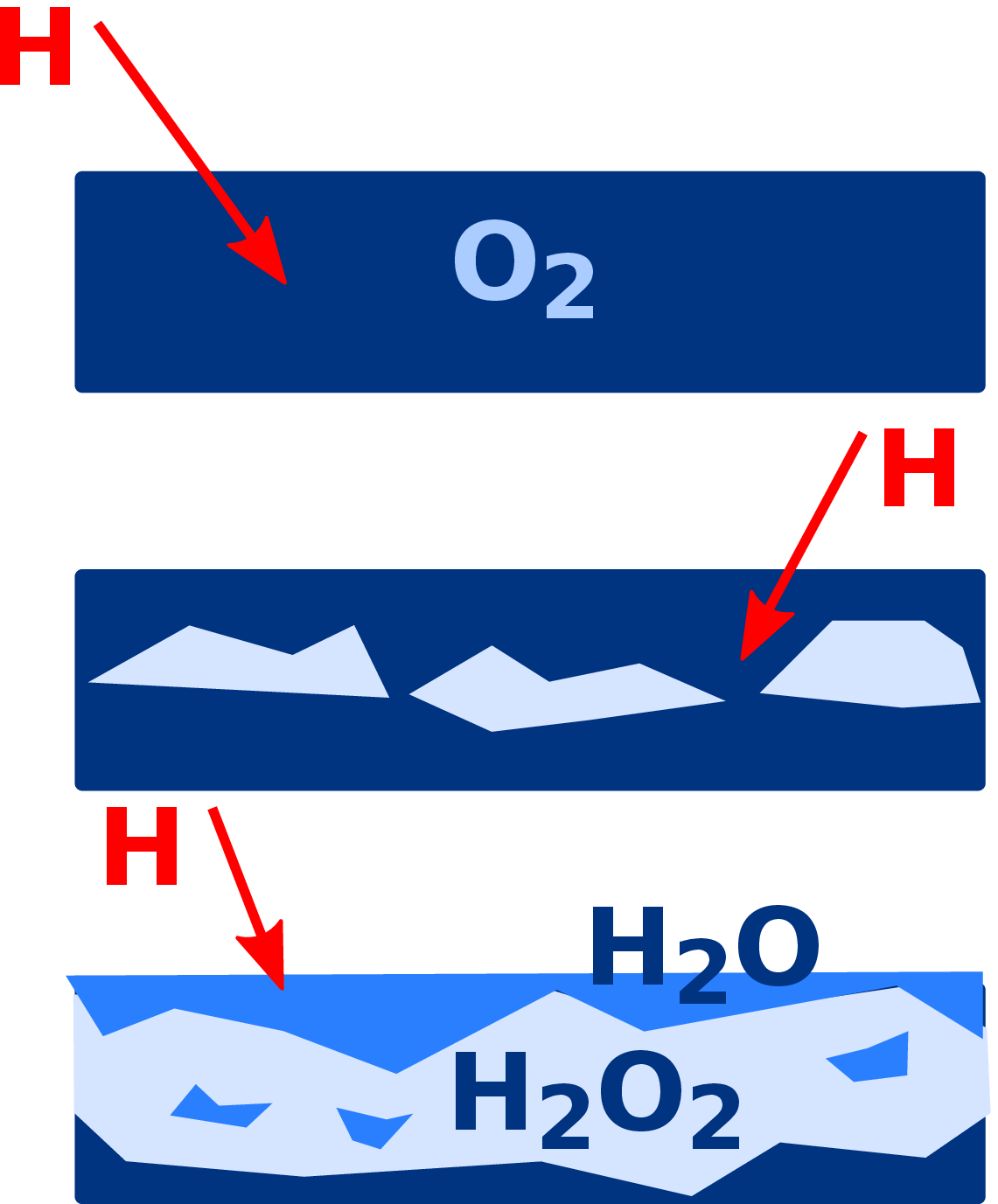}

\noindent
The penetration depth of cold H atoms into solid oxygen is affected by the competition between reaction and diffusion and increases with the temperature.
%Footnotes

\vspace{1ex}
\noindent
Due to a copyright agreement we are not allowed to publish the full paper on  arXiv.org. Please look \href{http://dx.doi.org/10.1039/C0CP00250J}{here} for the paper. We apologise for any inconvenience.

\footnotetext{\textit{$^{a}$}~Sackler Laboratory for Astrophysics,
Leiden Observatory, Leiden University, P. O. Box 9513, 2300 RA
Leiden, The Netherlands.} 
\footnotetext{\textit{$^{b}$}~Leiden
Observatory, Leiden University, P. O. Box 9513, 2300 RA Leiden, The
Netherlands.} 
\footnotetext{\textit{$^{c}$}~Max-Planck-Institut
f\"{u}r Extraterrestriche Physik, Giessenbachstrasse 1, D-85741
Garching, Germany.}

\footnotetext{\dag~Present address: LPMAA, Université Pierre et
Marie Curie, Paris}

\end{document}